# Dataset of raw and pre-processed speech signals, Mel Frequency Cepstral Coefficients of Speech and Heart Rate measurements

Mohammed Usman[1*] (SMIEEE), Zeeshan Ahmad[1], Mohd Wajid[2] (SMIEEE)

[1]King Khalid University, Abha, Saudi Arabia. Email: musman@ieee.org*, zayshan@kku.edu.sa
[2]Aligarh Muslim University, Aligarh, India. Email: wajidiitd@ieee.org

**Abstract**

Heart rate is an important vital sign used in the diagnosis of many medical conditions. Conventionally, heart rate is measured using a medical device such as pulse oxymeter. Physiological parameters such as heart rate bear a correlation to speech characteristics of an individual. Hence, there is a possibility to measure heart rate from speech signals using machine learning and deep learning, which would also allow non-invasive, non contact based and remote monitoring of patients. However, to design such a scheme and verify its accuracy, it is necessary to collect speech recordings along with heart rates measured using a medical device, simultaneously during the recording. This article provides a dataset as well as the procedure used to create the dataset which could be used to facilitate research in developing techniques to estimate heart rate accurately by observing speech signal.

## I. Introduction

Speech characteristics of an individual are related to physiological as well as emotional conditions of the speaker [1]. It should therefore be possible to determine physiological parameters or emotional state based on the speech of the individual. With the advent of machine learning (ML), such applications have the potential to becoming a reality. However, to develop, tune and optimize ML models that can accurately estimate a certain physiological parameter or emotional state, it is necessary to have data obtained using conventional methods. Lack of data or its scant availability hinders the research and development of such useful and novel applications. This article addresses this issue and provides a data set which can be used to develop ML models to estimate one of the important physiological parameter - heart rate from speech signals. Early work on measuring heart rate from speech is available in [2]–[4]. The equipment used and the methodology followed to create the data set is described in detail which will facilitate research in this direction. The complete specifications of the dataset discussed in this article is given in table 1. The rest of the paper is organized as follows. The value of the data in terms of its usefulness, scope, beneficiaries and applications is discussed in section II. Description and formats of the data in the dataset is provided in section III. In section IV, the materials and methods used to obtain the data, both raw and processed are described in detail. Finally, conclusions are drawn in section V along with a note on future work.

Table 1. Specifications Table

| | |
|---|---|
| **Subject** | Signal Processing |
| **Specific subject area** | Speech signal processing with machine learning |
| **Type of data** | Table, spreadsheet (.xlsx, .csv), speech recordings (.wav) |
| **How data were acquired** | - Speech recordings made using Logitech H540 microphone with noise cancellation in a quiet office room interfaced with Matlab.<br>- Heart rate measured concurrently with speech recording using a pulse oxymeter (CONTEC Model No. CMS50DL) through a microcontroller interface with the PC.<br>- Mel Frequency Cepstral coefficients(MFCC) are obtained by processing the recorded speech signals using a Matlab script. |
| **Data format** | Raw, Analyzed, Filtered |
| **Parameters for data collection** | - All volunteers are made to utter the same sentence and the recording is made at a sampling rate of 16000 samples per second.<br>- Depth of quantization is set to 16 bits per sample resulting in speech recordings at 256 kbps.<br>- Length of each recording is fixed to 5 seconds resulting in 80000 samples per recording<br>- Voice Activity Detection (VAD) and noise filtering is applied to remove 'silence' intervals and reduce noise in the recorded speech.<br>- MFCC data is computed for each speech recording by framing the speech samples using a fixed length Hamming window and a fixed overlap between adjacent windows.<br>- Heart rate is measured under resting condition, called the resting heart rate |
| **Description of data collection** | Speech and heart rate data measurements were collected using a microphone and pulse oxymeter respectively in a quiet office. MFCC data was computed by processing the recorded speech using a script in Matlab. |
| **Data source location** | The data described in this article was collected at<br>Institution: King Khalid University<br>City/Town/Region: Abha<br>Country: Saudi Arabia |
| **Data accessibility** | The complete data set which includes raw uncompressed speech recording (.wav), processed speech with VAD and noise removal (.wav), heart rate measurements (.xlsx) and MFCC data (.xlsx, csv) are available on Google Drive, which can be shared upon request. |

II. **Value of the Data**
- The data described and provided in this article will be extremely useful to develop novel techniques to measure heart rate from speech signals and facilitate research in this direction.

- Researchers, technologists, scientists and developers can use this data to develop and test machine learning algorithms that can estimate heart rate accurately from speech signals

- The data can be used in existing machine learning algorithms to determine their accuracy, tune the models to improve accuracy and also to develop new algorithms to achieve better accuracy between the estimated and measured heart rate values.

- While there are several speech datasets or corpuses available in the public domain, there is no dataset publicly available that has speech recordings along with heart rate measurements. Such a dataset is absolutely necessary to apply machine learning techniques to measure heart rate from speech.

- In addition to the recorded and measured data, processed data i.e. MFCC is also generated which can be used as 'features' for machine learning models.

- The feature dataset is significantly larger than the raw dataset which makes its particularly very useful for training and testing machine learning models.

III. **Description and format of data**

The dataset consists of a collection of four different types of data
- Raw speech files (in .wav format)
  The raw speech files contain the unprocessed, uncompressed speech recordings of 85 volunteers in .wav format.
- Heart rate measurements (as spreadsheet .xlsx format) containing measured heart rate and age of the volunteers.
  Heart rate measurements are made concurrently during the speech recording using a pulse oxymeter which is a conventional medical device to measure heart rate. This data is available in a spreadsheet (.xlsx format) which contains the measured heart rate, age and volunteer number in serial order. The identity of the volunteers is not provided in the dataset for compliance with 'data protection' regulations.
- Pre-processed speech files (in .wav format)
  The raw speech files are pre-processed to remove silence intervals ,pauses, DC offset and also to reduce noise resulting in speech files containing voiced activity with improved signal to noise ratio.
- Speech features files (MFCC) (in both .xlsx and .csv formats)
  Features are extracted from the pre-processed speech signals. MFCC is chosen as the feature extraction technique as it captures the spectral features along with their temporal variations. Features in several spectral bands are obtained for each speech signal, producing a manifold increase in the feature data set. The speech features data is available as both .xlsx and .csv formats

## IV.  Materials and Methods

### A.  Raw speech data and resting heart rate measurement

Raw uncompressed speech signals are first recorded in a quiet office environment using a Logitech H540 headset microphone. While recording the speech signal, heart rate of the individual is also measured using a pulse oxymeter. The pulse oxymeter measures the pulse rate which is exactly equal to the heart rate [5]. The volunteers whose data is collected are in the age group of 20 - 45 years and a total of 85 volunteers participated in this work. Each volunteer is asked to read the sentence 'A quick brown fox jumped over the lazy dogs' which is recorded through the microphone interfaced to the PC via a Matlab script. The sampling rate and depth of quantization is set to $f_S$ = 16000 samples per second 5and $n$ = 16 bits per sample respectively. Thus the resulting audio bit rate is $R_b = nf_S$ = 256 kbps. These parameters are adequate to faithfully represent speech signals [6], [7].These settings are configurable in the Matlab script which also activates the microphone. All instructions are provided to the volunteers who are made to rest for 30 minutes before recording their data in order to collect data corresponding to resting heart rate. It is also ensured that the volunteers have not performed any physical activity during at least the previous two hours before arriving to the recording room. The volunteer once settled puts on the headset microphone and the pulse oxymeter is attached to the left hand index finger and turned on. The volunteer is prompted to press any key on the keyboard whenever ready. Upon pressing any key, which is done only after the pulse oxymeter has started to capture its measurement, the microphone is activated for a duration of 5 seconds during which the speech is recorded in 'stereo' format, which records across two channels. The readings from the pulse oxymeter, which is interfaced to the PC via a micro-controller are also captured during these 5 seconds and stored on the PC along with the recorded speech. The average value of the heart rate obtained during these 5 seconds is taken as the heart rate corresponding to the recorded speech. Each speech recording, is stored on the PC as uncompressed .WAV format and is labelled with the speakers heart rate, which is stored as a spreadsheet in .XLSX format. At a sampling rate of 16000 samples per second, each 5 second recording results in 80000 samples per channel. This data comprises the raw speech files along with heart rates of the individuals corresponding to the resting heart rate.

### B.  Pre-processed speech data

The recorded speech signal may have pauses/gaps called 'silence intervals' in the beginning, end or in between the words of the sentence. Any delay on the part of the volunteer in speaking after pressing a key to activate the microphone results in a silence interval at the beginning resulting in only background noise and/or breathing sounds. Similarly, if the volunteer completes reading the sentence before the completion of the 5 seconds duration, it results in a silence interval towards the end of the recording. Silence intervals between words depends on the speaking style of each volunteer, i.e. their pronunciation or elongation of words and pauses made while reading. The silence intervals in speech contain no useful information for the purpose of measuring heart rate from speech. Hence, the silence interval is removed by applying a voice activity detection (VAD) algorithm, which also removes noise and any DC offset which might have been introduced by the PC audio card [8]. The VAD algorithm is implemented in Matlab using the procedures described in [9]. The raw speech recorded in stereo format consists of two channels and VAD algorithm is applied only to Channel-1 data resulting in pre-processed speech recordings in mono-format. These are also stored in uncompressed .WAV format with audio bit rate of 256 kbps as in the case of raw speech files. The pre-processed speech files are also labelled with the heart rate of the speaker.

### C. Speech features data

To utilize machine learning models for the estimation of heart rate from speech, it is necessary to extract appropriate features from the speech signals. A notorious trait of machine learning models is 'Garbage in, Garbage out'. Hence, it is important to provide appropriate data as input to machine learning models in order to obtain a meaningful and useful output. The features that are appropriate for a particular machine learning application depend on the type of data itself and also on the application for which the data is being used. There are several different ways of extracting features from speech signals, both in time domain as well as frequency domain. Some of the popular feature extraction techniques of speech are linear predictive coding (LPC), perceptual linear prediction (PLP), linear prediction cepstral coefficients (LPCC), Mel frequency cepstral coefficients (MFCC), wavelet based techniques, principal component analysis (PCA) and relative spectra (RASTA) technique [10]. In this data set, feature extraction is performed by computing MFCC for the pre-processed speech data. MFCC is used because it can represent spectral details of speech signals along with temporal variations in the spectral details. A detailed description of MFCC computation procedure is given in [11], [12]. The MFCC data generated in this work uses a Hamming window of length 256 samples which corresponds to speech frame duration of 16 ms, with 50% overlap between adjacent windows. The spectrum of each 16 ms speech obtained by computing a 256 point FFT from which the energy spectral estimate for the frame is determined. Since the sampling rate is set to 16000, the spectral range of each frame extends from 0 Hz to 8000 Hz. This spectral range is divided into 20 logarithmically spaced triangular windows with 50% overlap between adjacent windows. These are called as 'Mel' bands which mimic human perception of sound and collectively called as 'Mel' filterbank. The spectral energy within each Mel window is computed for each 16 ms speech frame and then converted to logarithmic scale. These filter-bank energies exhibit a high degree of correlation due to overlap between adjacent bands. The correlation is removed by applying Discrete Cosine Transform (DCT), which is known to be an optimal de-correlation technique for logarithmic spectra of speech [13], [14]. Thus the extracted features are in a matrix form in which each row represents a Mel frequency window and each column represents a speech frame of 16 ms duration. Since the spectral range is divided into 20 'Mel' bands the feature matrix contains 20 rows. The number of speech frames and hence the number of columns in the feature matrix depends on the speaking style of each individual which has an effect on the length/duration of silence intervals in recorded speech. In general, the dimension of MFCC matrix is 20 x I where 'I' is the number of speech frames in the pre-processed speech. The MFCC matrices are stored on PC as a spreadsheet in .XLSX and .CSV formats. An illustration of MFCC features is shown in figure 1 for a sample in the dataset.

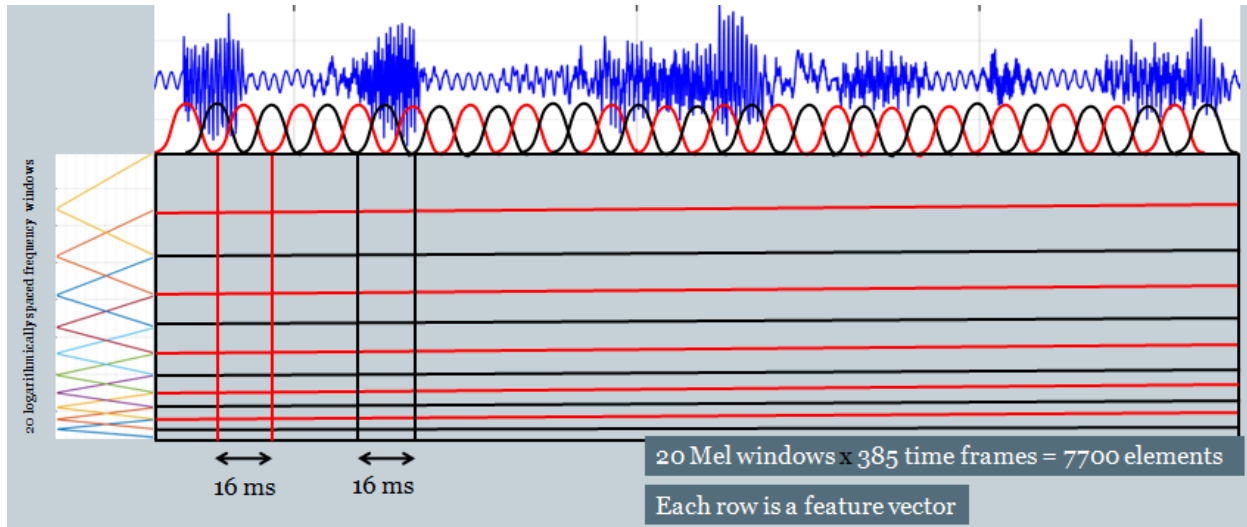

Figure 1. MFCC feature data for a sample in the dataset

## V. Conclusions

Speech characteristics of an individual can be used to measure certain physiological parameters such as heart rate of the individual. In order to develop machine learning models that can accurately estimate heart rate based on speech, it is required to acquire speech data along with measured heart rate. While machine learning is a promising tool for several applications, a major challenge is in the acquisition of reliable and accurate data which can be used to train and test machine learning models. In this article, a dataset is presented which can be used to estimate heart rate of an individual from his/her speech. Detailed methodology along with equipment and tools (both hardware and software) to create such a dataset has been described. The availability of such a dataset will facilitate research in the development, tuning and optimization of machine learning models that can estimate heart rate from speech which can have profound applications in medical diagnosis, treatment and care [15]. Future work shall consider expanding this data set by collecting more samples as well as measuring other parameters such as blood pressure and oxygen concentration in blood.